\documentstyle[12pt]{article}
\begin{document}

\hfill{LAEFF-94/10}

\hfill{WM-94-113}

\hfill{LA-UR-94-3996}

\hfill{November 1994}                      \vskip 12mm

\begin{center}

{\large
Gamma Ray Bursts, Neutron Star Quakes, and the Casimir Effect.}

\vskip 10mm

{C. E. Carlson}                            \vskip 2mm

{\it Physics Department, College of William and Mary,\\
         Williamsburg, VA 23187, USA}      \vskip 5mm

{T. Goldman}                               \vskip 2mm

{\it Theoretical Division, Los Alamos National Laboratory,

         NM 87545, USA}                    \vskip 5mm

{J. P\'erez--Mercader\footnote{Also at
        Instituto de Matem\'aticas y F\'{\i}sica Fundamental,

        CSIC, Serrano 119--123,
        E--28006 Madrid, Spain
               and
        Theoretical Division, Los Alamos National Laboratory,

        NM 87545, USA}}                    \vskip 2mm

{\it Laboratorio de Astrof\'{\i}sica Espacial y F\'{\i}sica
Fundamental,\\ Apartado 50727, E--28080 Madrid, Spain}

\end{center}                               \vskip 10mm

\begin{abstract}
We propose that the dynamic Casimir effect is a mechanism that
converts the energy of neutron starquakes into $\gamma$--rays.  This
mechanism efficiently produces photons from electromagnetic Casimir
energy released by the rapid motion of a dielectric medium into a
vacuum. Estimates based on the cutoff energy of the gamma ray bursts
and the volume involved in a starquake indicate that the total gamma
ray energy emission is consonant with observational requirements.
\end{abstract}


\newpage

Observationally, Gamma Ray Bursts (GRBs) are characterized by the
following generic parameters: (i) observed peak energy fluxes are of
the order $10^{-4}$ to $10^{-7}$ erg cm$^{-2}$
s$^{-1}$~\cite{kouveliotou,fenimore}, (ii) burst durations are
distributed according to a bimodal distribution with peaks at
$\approx
0.3$ sec and $\approx 25$ sec and median at  $\approx 10$ sec, and
(iii) observed photon energies run from a few times 10 keV out to 10
MeV~\cite{kouveliotou} and sometimes beyond~\cite{schneid92}. (The
larger figures can be taken as an effective momentum cutoff on the
Physics of the GRBs.)  In addition, GRBs are characterized by a
highly
non-thermal spectrum and show fluctuations over times on the order of
milliseconds~\cite{bhat92}.

For these GRBs to come from neutron stars and still fit within the
observed pattern of isotropy, the stars must be located in an
extended
galactic halo with a radius of the order of a few times $10^5$ ly.
(The
extended halo may be tied to the recent finding that supernova
explosions give rise to high velocity neutron stars, with the neutron
stars retained by Milky Way being distributed in a large isotropic
halo~\cite{ll94}.) Since BATSE sees down to fluences of about
$10^{-7}$
erg/cm$^{2}$, this translates into an emitted energy {\it at the
source} of~\cite{blaes}

\begin{equation}
E_{\rm Source} \approx  10^{41}\ {\rm erg}\times
          \left(D \over 3\times 10^5\ {\rm ly} \right)^2
          \left( F \over 10^{-7}\ {\rm erg/cm}^2 \right) .
\label{1}
\end{equation}

Any effect or theory that professes to explain GRBs must face and
explain this figure together with (i), (ii) and (iii) above.
Furthermore, since the energies are almost exclusively in the gamma
ray
range, one needs to identify a process that efficiently produces this
type of radiation.

One efficient process that could produce this kind of radiation is
the
Casimir effect. This well known effect \cite{casimir} consists on the
force appearing on a system of ``parallel interfaces in dielectric
media" due to the quantum nature of the electromagnetic force and its
associated quantum fluctuations. The Casimir force varies as the
fourth
power of the interface separation (unlike its classical antecedent,
the
electromagnetic force, which varies like $r^{-2}$), and therefore a
large amount of energy can be involved when the separation is very
small.  A system with conducting plates, where the plates are
accelerated to separate over a very small distance and then contract
(or
viceversa) can be a source of ``Casimir light" emitted as the system
relaxes \cite{schwingercasimirlight}. It is not difficult to imagine
that essentially this situation could occur at the surface of a
neutron
star when a starquake afflicts the star. {It may also happen in quark
stars, where there is a thin neutron $shell$ enveloping a quark
matter
core.}

In this letter, we will explore this scenario: a neutron star
undergoing a quake (we do not enter here into a discussion of the
mechanism giving rise to the starquake) and then relaxing. This gives
rise to a ``global'' Casimir effect and we estimate the total energy
released, the time scale for the phenomenon, and some general aspects
of the physics involucrated.

Previous applications of the Casimir effect to study little
understood
physical phenomena, include the series of papers (some of them
appearing at about the time of his death) written by the late J.
Schwinger to understand the phenomenon of
sonoluminescence~\cite{physicstoday}. Here we extend some of his
ideas
to the case of a quaking spherical neutron star. In the end, by
analogy, we infer that sonoluminescence in bubbles of liquid
material,
such as water, may very well provide us with a terrestrial,
laboratory
scale model for GRBs in a neutron star.

Typically, a neutron star quake deforms the surface of the star by a
$\Delta R$ of anywhere between $O(1\ {\rm cm})$  in PSR 0540 -- 69 to
$O(1\ \mu{\rm m})$ in a small Crab glitch~\cite{colgate}. ({The scale
of the deformation can be inferred from the change in angular
momentum
produced by the quake and then $assuming$ that the glitches observed
in
the periods of the pulsar are due to the starquake.}) Here $R$ is the
radius of the star, typically of the order of 10 km.  Assuming this
change is uniform, a vacuum shell may be formed by a separation
between
the stellar (neutron) core and the surrounding (Iron) surface; the
corresponding Casimir energy, given by

\begin{equation}
E_{Casimir} = \frac{\hbar c}{12 \pi}\frac{3}{4 \pi} \cdot V \cdot K^4
\cdot
\left(1- \epsilon^{-1/2} \right) \ ,
\label{3}
\end{equation}

\noindent
undergoes a change proportional to $\Delta R$. In Eq.(\ref{3}), $V$
denotes the volume of the star, $K$ is the momentum cutoff for the
radiation emitted on relaxation, and $\epsilon$ is the dielectric
constant for the medium.  Since a neutron star is close to a metal in
its electrical properties, we will take $\epsilon_{Neutron \ Star}
\rightarrow \infty$ for frequencies corresponding to photon momenta
below the cutoff.  At 100\% efficiency for the conversion of Casimir
energy into light, for a spherical neutron star undergoing a $\Delta
R
{\rm (cm)}$ starquake, the released Casimir energy is

\begin{equation}
\Delta E_{Cas.} = 15 \times \left[\frac{K {\rm (MeV/c)}}
                    {10\ {\rm MeV/c}}\right]^4
\times
\left[\frac{\Delta R {\rm (cm)} }{1\ {\rm cm}}\right] \times 10^{40}
{\rm erg}
\label{4}
\end{equation}

We see explicitly from this expression that (1) the Casimir energy
scales as the fourth power of the momentum cutoff and the first power
of the change in radius; thus, a very small change in the cutoff
scale
has a large effect. And (2) that the larger portion of the energy is
radiated in wavenumbers close to the cutoff. Therefore it does not
have
the same properties as a blackbody radiation, a point which can be
used
to test the validity of the present model when enough data are
accumulated. Of course, the flat specturm produced (up to the cutoff)
may well be attenuated at the higher energies by scattering on star
material or material in the surrounding system.

In the model we are describing here, the physics of the cutoff is
related to the physics of the collapse of the shell formed between
core
and surrounding surface, and is cognate to the effective,
differential
collapse velocity $\Omega$, by

\begin{equation}
\Omega^2 = \frac{\hbar c}{12 \pi^2}\frac{K^4}{\rho_0}
\left(1-\epsilon^{-1/2}\right)
\label{5}
\end{equation}

\noindent where $\rho_0$ is the density of the material.  This is the
speed that a shell of collapsing material would reach if $all$ its
Casimir energy were transformed into kinetic energy.

We consider two extreme values for $\Omega$: a lower limit obtained
by
assuming that the shell is free falling in the gravitational field of
the neutron star, and an upper limit obtained by assuming that the
shell falls with the speed of light. The two limiting values are
easily
computed from Eq.(\ref{5}), which gives $K_{lower}= 14$ MeV and
$K_{upper}= 400$ MeV.

Using these values and Eq.~(\ref{4}), we see that the total energy
released in each case (assuming 100\% energy conversion efficiency)
is

\begin{equation}
\Delta E_{Casimir}^{lower} = 57.6 \times
\left[\frac{\Delta R {\rm (cm)} }{1 {\rm cm}}\right] \times 10^{40}
{\rm erg}
\label{6}
\end{equation}

\noindent
and for the upper limit we get

\begin{equation}
\Delta E_{Casimir}^{upper} = 38.4 \times 10^{6} \times
\left[\frac{\Delta R {\rm (cm)} }{1\ {\rm cm}}\right] \times 10^{40}
{\rm erg}
\label{7}
\end{equation}

These two results clearly demonstrate that the two observational
values
for the total energy quoted above can be amply met by the release of
Casimir energy in the process of a starquake in a neutron star.  The
Casimir ``limiting speed" $\Omega$ given by Eq. ({\ref{5}}) that
results for a cutoff of 100 MeV can be readily calculated to be
$4.9\times 10^{8}$ cm/s, a little less than two percent the speed of
light.

The burst time observed on Earth is the time that the signal at this
speed takes to propagate around ({\it half}) the surface of the
neutron
star, and this is

\begin{equation}
T_{Total}= 6.5 \times 10^{-3} {\left[ \frac{100\ {\rm MeV/c}}{K{\rm
        (MeV/c)}}\right]}^2\ {\rm sec}
\label{8}
\end{equation}

\noindent in excellent agreement with the observational data for the
shorter bursts. Of course, there is no reason, in principle, why the
quake should not reverse its course across the star surface and so
extend the burst time.

Let us address the question of overall mechanism efficiency. This can
be estimated by comparing the change in the gravitational energy of
the
quaking star to the energy converted into gamma rays by the Casimir
effect as described here. Using Eq.(\ref{4}) and with Newtonian
gravity
and mechanics (justified because of the low $\Omega$) we have that
the
efficiency, $\eta$, required by the Casimir mechanism is given by

\begin{equation}
\eta = \frac{E_{Casimir}}{E_{Starquake}} \approx 2.5 \times
10^{-7}\left[\frac{K{\rm (MeV/c)}}{10\ {\rm MeV/c}}\right]^4
\label{9}
\end{equation}

\noindent that is, for a 100 MeV/c cutoff the efficiency need not be
any higher than about 0.25\%.

We close with a number of remarks.

The starquake-Casimir mechanism we have presented is not cataclysmic.
The neutron star survives and there could be repetitions of GRBs from
the same source on a time scale of years.  Our discussion uses
spherically symmetric starquakes for the sake of simplicity. This is
unlikely to be the case in practice, and so repetition---or
coincidence
with a pulsar glitch~\cite{hartmann}---would require also that the
starquake be in a region of the star facing Earth.

A cataclysmic version of a collapse-Casimir mechanism may be
possible,
and may be relevant if evidence that GRBs emanate from cosmological
distances becomes compelling.  For example, if a neutron star
accretes
mass from a nearby star and passes over the Chandrasekhar limit, it
will collapse into a black hole.  
vacuum does not give Casimir 
If
there is any bounce, or sudden outflow of dense material as a result
of
collapse, energetic Casimir radiation can ensue.

Details of the burster length and time structure depend on details of
the infall, reverberation, and lateral spread of a starquake. Further
work on this as well as on equation of state effects, effects of
finite
permittivity, and {\it ab initio} calculation of the cutoff wave
number
is under study.  The reason for the existence of two classes of GRBs,
if indeed the two classes are really distinct, is also open, although
we expect it is related  to the preceding list.

In summary, we have presented the dynamic Casimir effect as a
mechanism
that could efficiently explain the conversion of starquake energy
into
gamma rays. The total energy converted into gammas depends on the
gamma
ray cutoff energy and the volume of material involved in the
starquake.  Furthermore, our estimates are in encouraging agreement
with observation.

\medskip {\bf Acknowledgements} {We thank Stirling Colgate,
Stuart Bowyer, and Alberto Castro--Tirado for discussions on
GRBs, and G. Luther for his discussion of sonoluminescence.  CEC
and JPM thank the members of the Theoretical Division at LANL
for their hospitality during the initial phase of this work.

CEC also thanks the NSF-USA for support under Grant PHY-9306141.}

\end{document}